\documentclass[preprint,aps,12pt,preprintnumbers,eqsecnum,nofootinbib]{revtex4}
\usepackage{xcolor}
\usepackage{graphicx}
\usepackage{subfigure}
\usepackage{epstopdf} 
\usepackage{braket}
\usepackage{amssymb,amsmath,amsfonts,comment,latexsym,graphicx,epstopdf,float}
\usepackage{color}
\usepackage{fancyvrb}
\usepackage{chngcntr}
\usepackage{etoolbox}
\usepackage{bbm,epsfig}
\usepackage{array}
\usepackage{slashed}

\usepackage[colorlinks=true,citecolor=blue,linkcolor=red,filecolor=cyan,urlcolor=magenta]{hyperref}

\newcommand{\be}{\begin{equation}}
\newcommand{\ee}{\end{equation}}
\newcommand{\ba}{\begin{eqnarray}}
\newcommand{\ea}{\end{eqnarray}}

\newcommand{\grts}{\raise.3ex\hbox{$>$\kern-.75em\lower1ex\hbox{$\sim$}}}
\newcommand{\lets}{\raise.3ex\hbox{$<$\kern-.75em\lower1ex\hbox{$\sim$}}}

\newcommand{\dd}{\text{d}}

{\catcode`\|=\active\gdef\Braket#1{\left<\mathcode`\|"8000\let|\bravert 
{#1}\right>}}

\def\bravert{\egroup\,\vrule\,\bgroup}

\usepackage{subfigure}

\usepackage{color}
\usepackage{amssymb,amsmath,amsfonts}
\usepackage{epstopdf} 
\usepackage{braket}

\begin{document}
%
%
\title{\vspace*{0.5in} 
Asymptotic nonlocality
\vskip 0.1in}
\author{Jens Boos}\email[]{jboos@wm.edu}
\author{Christopher D. Carone}\email[]{cdcaro@wm.edu}
\affiliation{High Energy Theory Group, Department of Physics,
William \& Mary, Williamsburg, VA 23187-8795, USA}

\date{\today}
%
%
\begin{abstract}
We construct a theory of real scalar fields that interpolates between two 
different theories: a Lee--Wick theory with $N$ propagator poles, including $N-1$
Lee--Wick partners, and a nonlocal infinite-derivative theory with kinetic terms modified by an entire function of derivatives with only one propagator pole. 
Since the latter description arises when taking the $N\rightarrow\infty$ limit, we refer to the theory as ``asymptotically nonlocal.''  Introducing an 
auxiliary-field formulation of the theory allows one to recover either the higher-derivative form (for any $N$) or the Lee--Wick form of the Lagrangian, 
depending on which auxiliary fields are integrated out. The effective scale that regulates quadratic divergences in the large-$N$ theory is the would-be 
nonlocal scale, which can be hierarchically lower than the mass of the lightest Lee--Wick resonance. We comment on the possible utility of this construction in addressing 
the hierarchy problem.
\end{abstract}

\maketitle

\section{Introduction}\label{sec:intro}
When the Higgs boson was discovered at the Large Hadron Collider (LHC) in 
2012 at a mass of 125 GeV, an immediate question was: why is it so light compared to the Planck scale at $10^{19}$~GeV? The standard model, in its 
present form, does not provide a mechanism that protects the Higgs mass from picking up large radiative corrections from Planck-scale physics. And 
if there was no such mechanism, then the observed Higgs mass had to be the result of an incredibly fine-tuned almost-cancellation between the bare 
mass and its corrections.

This has been dubbed the ``hierarchy problem,'' and while there have been 
many proposed solutions, in this paper we would like to focus on the Lee--Wick approach \cite{LeeWick:1969}. The Lee--Wick standard model \cite{Grinstein:2007mp} provides an inherent mechanism to address the hierarchy problem: supplementing the spectrum of the standard model by TeV-scale Lee--Wick partner particles with wrong-sign kinetic and mass terms leads to a cancellation of quadratic divergences in scalar self-energies, with the 
precise spectrum of Lee--Wick resonances left for experiments to determine. These models have been shown to be unitary, provided certain prescriptions are adopted in momentum space to deal with the different pole structure \cite{Cutkosky:1969fq} (see also \cite{Anselmi:2017yux}), are causal at the macroscopic level \cite{Grinstein:2008bg}, have subsequently been generalized to include additional partner particles \cite{Carone:2008bs,Carone:2008iw}, and have garnered much attention in the phenomenological literature as well \cite{LWpheno}. However, to date, no such Lee--Wick partner particles have been found up to $\sqrt{s} \sim \mathcal{O}(10)\,\text{TeV}$ \cite{Zyla:2020zbs}, and while shifting these masses to higher, unprobed energies would in principle be permissible, it would no longer provide a convincing solution of the hierarchy problem.

The last decades have also seen the development of nonlocal field theories \cite{Efimov:1967,Krasnikov:1987,Kuzmin:1989,Tomboulis:1997gg,Modesto:2011kw,Biswas:2011ar}, in particular those without additional degrees of freedom, sometimes referred to as ``ghost-free'' theories. While these Lorentz-invariant nonlocal theories were originally intended as simple generalizations of Pauli--Villars regularization schemes \cite{Wataghin:1934ann,Pauli:1949zm,Pais:1950za} they have since been rediscovered in p-adic string theory and non-commutative geometry \cite{Witten:1986,Frampton:1988,Tseytlin:1995uq,Siegel:2003vt}. A large body of research has been devoted to a better understanding of \emph{ab initio} nonlocal theories, and recent work and applications include regular bouncing cosmologies \cite{Biswas:2005qr}, nonlocality in string theory \cite{Calcagni:2013eua,Calcagni:2014vxa}, regularization of the gravitational field via nonlocality \cite{Edholm:2016hbt,Boos:2018bxf,Giacchini:2018wlf}, as well as the role of the Wick rotation vis-\`a-vis unitarity and causality \cite{Carone:2016eyp,Tomboulis:2015esa,Shapiro:2015uxa,Briscese:2018oyx,Briscese:2021mob,Koshelev:2021orf}. For a review of infinite-derivative quantum field theory see Ref.~\cite{Buoninfante:2018mre}, and for additional references we refer the reader to Ref.~\cite{Boos:2020qgg}.

In nonlocal theories, the scale of nonlocality is set by a free parameter 
that we will call $M_{\rm nl}$; in ghost-free models, this scale does not 
correspond to 
the mass of a resonance as there are no corresponding poles in the propagator.  Given the current status of LHC searches for new particles, this is potentially a desirable property for a realistic nonlocal extension of the standard model.  One might expect that non-resonant effects on scattering 
amplitudes would allow $M_{\rm nl}$ to be as low as the typical, TeV-scale collider bounds on new physics that does not violate flavor, CP or 
Lorentz invariance.  In the present work, we develop a simple scalar toy model that interpolates between a Lee--Wick and a nonlocal theory, where the 
latter is obtained as the number of Lee--Wick resonances tends to infinity and the spectrum of heavy particles decouples.   What is interesting about this 
``asymptotically nonlocal" theory is that the would-be nonlocal scale, $M_{\rm nl}$, becomes the effective regulator of scalar loop diagrams, while 
remaining hierarchically lower than the mass of the lightest Lee--Wick resonance.  (For another proposal in which the effective scale of nonlocality is 
expected to be much lower than a cutoff scale see \cite{Buoninfante:2018gce}.)  In the case where the number of Lee--Wick resonances is finite but 
large, the hierarchy problem in this particular model is resolved even as 
the Lee--Wick resonances are decoupled from the low-energy effective theory.

It should be stressed that our goal is to construct specific Lee--Wick theories with $N$ propagator poles that have
the decoupling and divergence properties described above.  These higher-derivative theories are strictly local and avoid a pathology of the
nonlocal theory that they approach, namely, the inability to Wick rotate in Minkowski space leading to  a loss of 
unitarity~\cite{Carone:2016eyp,Tomboulis:2015esa,Shapiro:2015uxa,Briscese:2018oyx,Briscese:2021mob,Koshelev:2021orf}.
Our finite-$N$ Lee--Wick theories are treated as fundamental and are of interest in their own right; the significance of the  $N \rightarrow \infty$ 
limiting form is only in that it suggests an organizing principle that assures the emergence of a derived mass scale that 
regulates loop integrals, one that is hierarchically smaller than the lightest Lee--Wick partner mass.    Our approach is in some sense analogous to 
dimensional deconstruction~\cite{ArkaniHamed:2001ca,Hill:2000mu}, where interesting four-dimensional theories can be defined whose 
properties may be understood in terms of a five-dimensional limiting theory, even though one is never actually interested in taking or exactly 
reproducing the limit.  Just as a deconstructed model should not be thought of as a truncation of a five-dimensional theory for the purposes of 
approximation, the same perspective applies to the finite $N$ theories that we are about to present.

This paper is organized as follows: In Sec.~\ref{sec:framework}, we develop our basic framework of an auxiliary field theory, and prove that 
integrating out the auxiliary fields results in an effectively higher-derivative theory, and---in the limit of infinitely many auxiliary fields---results in a 
manifestly nonlocal theory; we refer to this as asymptotic nonlocality.  In Sec.~\ref{sec:lw}, we match this theory to a Lee--Wick theory, and as an application in Sec.~\ref{sec:qd} we compute the scalar self-energy at finite and infinite $N$, at one loop. We show for this simple model that asymptotic 
nonlocality can indeed separate the scale of the lightest Lee--Wick resonances from the scale that characterizes the convergence of loop diagrams.
We summarize our findings in Sec.~\ref{sec:conc} and suggest that no impediments exist to generalizing this framework to Abelian gauge theories, 
non-Abelian gauge theories, and, in principle, the entire standard model.

\section{Framework} \label{sec:framework} 
Consider the following (nongeneric) theory of  $N$ real scalar fields $\phi_j$, and $N-1$ real scalar fields $\chi_j$:
\begin{equation}
{\cal L}_N = -\frac{1}{2} \, \phi_1 \Box \phi_N - V(\phi_1) - \sum_{j=1}^{N-1} \chi_j \, \left[ \Box \phi_j - (\phi_{j+1}-\phi_j)/a_j^2\right] \,\,\, .
\label{eq:start}
\end{equation}
Here, the $a_j$ are $N-1$ constants with dimension of length, and we have 
set all the coefficients of the terms involving the 
$\chi_i$ to one, for simplicity. The signature of the metric is $(+,-,-,-)$ and $\Box \equiv \partial_\mu\partial^\mu$. Functionally integrating over 
the $\chi_i$ leads to the constraints\begin{equation}
\Box \phi_j -(\phi_{j+1}-\phi_j)/a_j^2=0 \,\,\, ,\,\,\,\,\,\, \mbox{ for }j=1 \dots N-1.
\label{eq:recursive}
\end{equation}
Equation \eqref{eq:recursive} is exact at the quantum level, since functionally integrating over the $\chi_j$, which appear linearly in the
Lagrangian, yields a functional delta function in the generating functional for the correlation functions of the theory.\footnote{ Note that 
taking the $a_j \rightarrow 0$ limit in Eq.~(\ref{eq:recursive}) transforms the difference $\phi_{j+1}-\phi_j$ into a derivative with respect to a 
new 
parameter; the approach of exchanging nonlocality for locality in a higher-dimensional space has been explored in 
Refs.~\cite{Calcagni:2007ef,Calcagni:2010ab}; see also \cite{Kan:2020ssy}.} This delta function 
allows one to perform the functional integration over $N-1$ of the $\phi_j$ fields, which we will take as those with $j=2\ldots N$; this yields a 
theory for $\phi_1$. It is instructive first to consider the case where the $a_j$ are all equal, {\em i.e.}, $a_j \equiv a$.
Then recursive application of Eq.~(\ref{eq:recursive}) implies
\begin{equation}
\phi_N \equiv \left(1+\frac{\ell^2 \Box}{N-1}\right)^{N-1} \phi_1 \,\,\, ,
\end{equation}
where $\ell^2 \equiv (N-1) \, a^2$.  Letting $N \rightarrow \infty$ while 
holding $\ell$ fixed one finds
\begin{equation}
\phi_\infty =  e^{\ell^2 \Box} \phi_1 \,\,\, ,
\end{equation}
such that the resulting Lagrangian takes the form
\begin{equation}
{\cal L}_\infty = -\frac{1}{2} \, \phi_1 \, \Box \, e^{\ell^2 \Box} \, \phi_1 - V(\phi_1) \,\, .
\label{eq:Linf}
\end{equation}
This is a familiar form found in the literature on nonlocal quantum field 
theories, with the scale of nonlocality fixed by the 
parameter $\ell$. It is important to note that the $\Box$ operator appearing in the exponential is not necessarily the best 
choice for a nonlocal theory defined in Minkowski spacetime due to time-dependent instabilities \cite{Frolov:2016xhq,Boos:2019fbu}.
However, our construction holds if one replaces the $\chi_j \,\Box \phi_j$-terms in Eq.~\eqref{eq:start} with $\chi_j \,{\cal D} \phi_j$,
where ${\cal D}$ can be any $\chi_j$-independent differential operator that one would like to appear in the exponential. \emph{Inter alia,}
this includes the Lorentz-violating possibility that ${\cal D}=\vec{\nabla}^2$, discussed in Ref.~\cite{Carone:2020zfk}.

The choice of a single $a_j \equiv a$ is convenient for the purpose of illustration, but it has the following drawback:  at finite $N$, 
the propagator of the theory,
\begin{equation}
D_F(p^2) = \frac{i}{p^2 \, (1-a^2 \, p^2)^{N-1}} \, ,
\label{eq:badprop}
\end{equation}
cannot be written as a sum over simple poles. [Here we have assumed, for simplicity, that there is no $\phi_1$ mass term in the 
potential $V(\phi_1)$.] Alternatively, Eq.~\eqref{eq:badprop} can be written as a sum over simple poles, but only in a limit where the 
wave function renormalization of all but one factor diverges. In what follows, we avoid this problem by assuming that the $a_j$ are
nondegenerate. The higher-derivative form of the theory is then given by
\begin{equation}
\label{eq:hd-lagrangian}
{\cal L}_N = - \frac{1}{2} \phi_1 \Box \left[\prod_{j=1}^{N-1} \left(1+\frac{\ell_j^2 \Box}{N-1} \right) \right] \phi_1 - V(\phi_1)  \,\,\, ,
\end{equation}
where $\ell_j^2 \equiv (N-1) \, a_j^2$, and where the propagator now takes the form
\begin{equation}
D_F(p^2) = \frac{i}{p^2} \, \prod_{j=1}^{N-1} \left(1-\frac{\ell_j^2 \, 
p^2}{N-1}\right)^{-1}\,\,\, .
\label{eq:nondegprop}
\end{equation}
One would expect that the same limiting form of the Lagrangian, Eq.~\eqref{eq:Linf}, is obtained when $N\rightarrow \infty$ and the $\ell_j$ 
approach a common, nonvanishing value.  This finite-$N$ theory provides a 
spectrum of nondegenerate states that alternate between 
ordinary particles and ghosts~\cite{Pais:1950za}, {\em i.e.}, particles whose propagators have wrong-sign residues.  This is what one would expect 
in Lee--Wick theories that have additional higher-derivative quadratic terms~\cite{Carone:2008iw} beyond those found in the Lee--Wick standard 
model~\cite{Grinstein:2007mp}, see also \cite{Krasnikov:1987}.  The physics of the theory at large but finite $N$ is not
strictly equivalent to that of the limit $N \rightarrow \infty$ where additional pathologies arise (see Sec.~\ref{sec:intro}).     
Nevertheless, by working with local theories at finite $N$, we inherit a desirable feature of the limiting theory, namely a derived scale that regulates
loop integrals and that can be hierarchically smaller than the mass scale of the lightest Lee--Wick partner, $a_1^{-1}$. We show this explicitly later.

\section{Lee--Wick Form of the Theory} \label{sec:lw}
At finite $N$, the theory described in the previous section predicts $N$ physical poles, as can be seen from Eq.~(\ref{eq:nondegprop}), which 
follows immediately from the higher-derivative form of the theory for $\phi_1$.  We obtained this higher-derivative form by integrating out 
$N-1$ of the $2\,N-1$ scalar fields in Eq.~(\ref{eq:start}).  The theory can also be written in a form without higher-derivatives, in which there is 
a separate field corresponding to each of the $N$ poles, with canonical quadratic terms up to overall signs; in the case of the Lee--Wick 
standard model, this has been called the Lee--Wick form of the theory~\cite{Grinstein:2007mp}, a terminology we adopt here. The same form can be
achieved in the present model by integrating out a different choice of $N-1$ fields that appear linearly in Eq.~(\ref{eq:start}). Let us illustrate
this in the present section, in the cases where $N=2$ and $N=3$, and comment on generalizations that are required for arbitrary $N$ at the end.

\subsection{Case $N=2$}
For the case of $N=2$, let us represent the field basis by $\Psi \equiv 
(\phi_1,\chi_1,\phi_2)$.  The quadratic terms in Eq.~\eqref{eq:start} may 
then be written as
\begin{equation}
{\cal L}_\text{quad} = - \frac{1}{2} \, \Psi^T (K\,  \Box + M) \Psi \,\,\, ,
\end{equation}
where the matrices $K$ and $M$ are given by
\begin{equation}
K = \left(\begin{array}{ccc} 0 & \,\, 1 \,\,  & 1/2 \\ 1 & 0 & 0 \\ 1/2 
& 0 & 0 \end{array} \right) \,\,\,\,\, \mbox{ and } \,\,\,\,\,
M= \left(\begin{array}{ccc} 0 & \,\, 1/a_1^2 \,\, & 0 \\ 1/a_1^2 & 0 & -1/a_1^2 \\ 0 & -1/a_1^2 & 0 \end{array} \right) \,\,\, .
\end{equation} 
We seek an invertible matrix $S_2$ such that $K_0=S_2^T K S_2$ and $M_0=S_2^T M S_2$ are simultaneously diagonal (or as close to it as possible).   With the 
choice
\begin{equation}
S_2 = \left(\begin{array}{ccc} 1 & -1 & 0 \\ 0 & \,\, 1/2 & \xi \\ 1 & \,\, 0 & -2\, \xi \end{array} \right) \,\,\ ,
\label{eq:S2}
\end{equation}
where $\xi$ is an arbitrary nonvanishing real parameter, one finds
\begin{equation}
K_0 = \left(\begin{array}{ccc}  1 & \,\, 0 \,\,  & 0 \\ 0 & -1 & 0 \\ 0 
& 0 & 0 \end{array} \right) \,\,\,\,\, \mbox{ and } \,\,\,\,\,
M_0= \left(\begin{array}{ccc}  0 & \,\, 0 \,\, & 0 \\ 0 & -1/a_1^2 & 0 \\ 0 & 0 & 4\, \xi^2/a_1^2  \end{array} \right) \,\,\, .
\label{eq:dforms}
\end{equation} 
A number of comments are in order. A combination of unitary rotations and 
rescalings are sufficient to put the kinetic matrix in the form
shown in Eq.~(\ref{eq:dforms}). The remaining transformations needed to diagonalize $M$ must preserve the form of $K_0$ [for example, an SO(1,1) rotation
in the upper-left two-by-two block]; the matrix $S_2$ encodes all of these transformations. The mass of the $\Psi_3$ field appears to be dependent 
on
an arbitrary parameter $\xi$, but $K_0$ reveals that this linear combination of the original fields has no kinetic term and hence does not correspond to a physical
degree of freedom. As before, integrating out $N-1$ fields (in this case, 
the one field $\Psi_3$, whose functional integral can be performed exactly) leaves one
with a theory with two physical degrees of freedom, an ordinary particle (with vanishing mass by our initial choice), and a Lee--Wick partner with 
mass $1/a_1^2$. 
This matches our expectation from the higher-derivative form of the theory, including the correct locations of the poles. Note that the form of $S_2$ ensures that the potential $V(\phi_1)$ has no dependence 
on the auxiliary field $\Psi_3$.

\subsection{Case $N=3$}
Let us further strengthen our intuition for extracting the Lee--Wick form 
of the theory by considering the theory with $N=3$. In this case, the kinetic and mass matrices are
\begin{equation}
K = \left(\begin{array}{ccccc} 0 & \,\, 1 \,\,  & 0 & 0 & 1/2 \\ 1 & 0 & 0 & 0 & 0 \\ 
0 & 0 & 0 & \,\, 1 \,\, & 0 \\
0 & 0 & \,\, 1 \,\, & 0 & 0 \\
1/2 & 0 &0&0 & 0 \end{array} \right) \,\,\,\, \mbox{ and } \,\,\,\,
M= \left(\begin{array}{ccccc} 0 & \,\, 1/a_1^2 \,\, & 0  & 0 &0\\ 1/a_1^2 & 0 & -1/a_1^2 &0 &0 \\ 0 & -1/a_1^2 & 0 & 1/a_2^2 & 0 \\
0 & 0 & 1/a_2^2 & 0  & -1/a_2^2 \\
0 & 0 & 0 & -1/a_2^2 & 0  \end{array} \right) \,\,\, ,
\end{equation} 
in the field basis $\Psi \equiv (\phi_1,\chi_1,\phi_2,\chi_2,\phi_3)$.
Again, we seek an invertible matrix (called $S_3$) that performs a simultaneous diagonalization. In this case, only a block diagonalization is possible. With the
choice
\begin{equation}
S_3 = \left(\begin{array}{ccccc} 1 \,\, & \,\, -\sqrt{\frac{y}{y-1}} \,\, & \,\, -\frac{1}{\sqrt{y-1}} \,\,& 0 & 0 \\
0 & \frac{1}{2}\sqrt{\frac{y-1}{y}} & 0 & \frac{y-1}{4 \sqrt{y}} & -\xi/2 
\\
1 & 0 & \sqrt{y-1} & -\sqrt{y} & 0 \\
0 & \frac{1}{2 \sqrt{y}} \frac{1}{\sqrt{y-1}} & \frac{1}{2} \frac{1}{\sqrt{y-1}}& \frac{1}{2 \sqrt{y}} & 0 \\
1 & 0 & 0 & -\frac{1+y}{2\sqrt{y}} & \xi \end{array} \right) \,\,\, ,
\label{eq:S3}
\end{equation}
 where $y\equiv a_1^2/a_2^2 > 1$ and $\xi$ is again an arbitrary nonvanishing real parameter, one finds that $K_0=S_3^T K S_3$ and $M_0=S_3^T M S_3$
are given by
\begin{equation}
K_0 = \left(\begin{array}{ccccc}  1 & \,\, 0 \,\,  & 0 & 0& 0\\ 0 & -1 & 0 &0 &0 \\
0 & 0 &1 & 0 & 0 \\
0 & 0 &0 & -1 & 0 \\
0 & 0 & 0  & 0 & 0 \end{array} \right) \,\,\,\,\, \mbox{ and } \,\,\,\,\,
M_0= \left(\begin{array}{ccccc}  0 & \,\, 0 \,\, & 0 & 0 & 0\\ 0 & -1/a_1^2 & 0 & 0 & 0\\ 
0 & 0 &1/a_2^2 & 0 & 0 \\
0 & 0 & 0 & 0 & -\frac{\xi}{a_1 a_2} \\
0 & 0 & 0 &  -\frac{\xi}{a_1 a_2} & 0 
\end{array} \right) \,\,\, .
\label{eq:d3forms}
\end{equation} 
The two-by-two block in the lower-right of $K_0$ and $M_0$ cannot be simultaneously diagonalized, and also depends on an undetermined parameter $\xi$.
Both are indications that this block corresponds to degrees of freedom that are unphysical.  The field $\Psi_5$ only appears linearly through $M_0$ and it can be
functionally integrated, leaving a functional delta function that sets $\Psi_4$ equal to zero.  As before, integrating out $N-1$ fields (in this case $\Psi_4$ and
$\Psi_5$) we are left with fields that represent the physical degrees of freedom expected in the Lee--Wick form of the theory, {\em i.e.}, two Lee--Wick partners, one 
of which is a ghost.   The mass spectrum again matches our expectation from the higher-derivative form of the theory (see also Ref.~\cite{Carone:2008iw}).
In analogy to our previous results, the form of $S_3$ ensures that the potential $V(\phi_1)$ has no dependence on the auxiliary fields $\Psi_4$ and $\Psi_5$.

\subsection{General structure}
Based on our gained intuition, let us now make some more general comments 
applicable to general $N$. In our loop calculation discussed in the next section, we will
focus on the mass renormalization of the field $\Psi_1$. As it turns out, 
it is most easily determined by working in the higher-derivative form of the theory for $\phi_1$, and
by extracting the part of this field on external lines that corresponds to the lightest physical mass eigenstate $\Psi_1$. Based on Eqs.~\eqref{eq:S2} and \eqref{eq:S3}
we see that this part is easily identified as having unit coefficient,  {\em i.e.}, $\phi_1 = \Psi_1 + \cdots$.  This fact is true for arbitrary 
$N$, which can be
verified by considering the partial fraction decomposition of the higher-derivative propagator for nondegenerate $a_j$,
\begin{equation}
\frac{i}{p^2} \prod_{j=1}^{N-1} \left(1-a_j^2 \, p^2 \right)^{-1} = \frac{i}{p^2}\, b_0  + \sum_{k=1}^{N-1} \, \frac{i}{p^2 - 1/a_k^2} \, b_k \,\,\, ,
\label{eq:pfd1}
\end{equation}
which holds algebraically for
\begin{equation}
b_0=1 \,\,\,\,\, \mbox{ and } \,\,\,\,\, 
b_k = \left\{ \begin{array}{cr} - 1 & \mbox{ for }N=2 \\[3pt]
 -\prod\limits_{\substack{j=1\\j\not=k}}^{N-1} \left(1-a_j^2/a_k^2 \right)^{-1} & \mbox{ for } N \geq 3 \end{array} \right.
 \label{eq:pfd}
\end{equation} 
The wave function renormalization of the lightest mass eigenstate, {\em i.e.}, $b_0=1$, ensures that a self-energy diagram with two external $\phi_1$ lines and one with
two external $\Psi_1$ lines are related by a relative factor of $+1$. We also note in passing that the $b_k$ above are in agreement with the square of the elements of the 
first rows of $S_2$ and $S_3$, respectively, providing a useful consistency check.

With the basic structures in the finite-$N$ theory understood, let us now 
apply these techniques to study the quadratic divergences in such a theory.

\section{Quadratic Divergences}\label{sec:qd}

As we remarked in the previous section, the field $\Psi_1$ represents the 
lightest physical state, and it is interesting to consider its mass renormalization, generated by a 
suitable self-interaction $V(\phi_1)$. We shall work in the higher-derivative theory \eqref{eq:hd-lagrangian} where we choose a quartic self-interaction
\begin{align}
V(\phi_1) = \frac{\lambda}{4!} \phi_1^4 \, .
\end{align}
At one-loop, the self-energy for the $\phi_1$-field, $M^2(k^2)$, is given 
by
\begin{align}
-i M^2(k^2) = -i\, \frac{\lambda}{2} \int \frac{\dd^4 p}{(2\pi)^4} \, D_F(p^2) = \frac{\lambda}{2} \int \frac{\dd^4 p}{(2\pi)^4} \, \frac{1}{p^2} \prod\limits_{j=1}^{N-1} \big( 1 - a_j^2 p^2 \big)^{-1} \, ,
\label{eq:selfen}
\end{align}
where $k$ is the external line momentum and we have inserted the higher-derivative propagator \eqref{eq:nondegprop}.  The two 
external $\phi_1$ lines can be reexpressed in terms of the tree-level mass eigenstates $\Psi_j$, using the type of decomposition described in the 
previous section.  We focus here on the case of two external $\Psi_1$ lines, which does not alter any numerical factors in
Eq.~(\ref{eq:selfen}); mixing effects induced at one-loop between $\Psi_1$ and $\Psi_j$, for $j>1$, become negligible in the 
limit of interest, {\em i.e.}, where the tree-level masses are taken such 
that latter states are decoupled.  It is our goal to compute this 
self-energy for finite $N$, as well as the asymptotic form of this result 
as $N \rightarrow \infty$.  In what follows, we will employ the notation 
$m_j^2 = 1/a_j^2$ and assume that all masses are nondegenerate. As shown in the previous section, these masses correspond to the poles of the 
higher-derivative propagator, and match those of the Lee--Wick theory as well, in both cases appearing with residues of 
alternating sign.

For $N$ that is finite but not large, we will see that the self-energy scale is set by the smallest of the Lee--Wick partner masses $m_j^2$, as 
expected.  This means that protecting the lightest scalar mass from large 
radiative corrections would not allow us to decouple the spectrum
of Lee--Wick partner states. A different conclusion is reached 
when $N$ becomes large, as we will show by constructing an exactly solvable, nondegenerate mass-parametrization that allows us to perform the $N\rightarrow\infty$ limit analytically. In this case, we will prove that the one-loop contribution to the scalar mass squared is set by the nonlocal 
scale $M^2_\text{nl} \equiv 1/\ell^2$, where
\begin{align}
\label{eq:hierarchical-separation}
M^2_\text{nl} \sim {\cal O}\left(\frac{m_1^2}{N}\right) \, ,
\end{align}
where $m_1$ is the mass of the lightest Lee--Wick resonance.  This provides a mechanism of achieving a hierarchical separation of scales.

\subsection{Partial Fraction Method at Finite $N$} \label{subsec:a}
We first note that the higher-derivative propagator $D_F(p^2)$ in Eq.~(\ref{eq:selfen}) has the partial fraction decomposition given in
Eq.~(\ref{eq:pfd1}).  Each term of the right-hand-side is of a form suitable for Wick rotation, without any unusual prescriptions required in
Lee--Wick theories for more complicated denominators~\cite{Cutkosky:1969fq}.  As a result, we can do the same on the left-hand-side and 
study the self-energy of interest by working in Euclidean space from the start ($p^0 \rightarrow ip_E^0$ and $p^2 \rightarrow -p_E^2$).  Hence, 
\begin{equation}
-i M^2(k^2) = - \frac{ i \lambda}{2} \int \frac{\dd^4 p_E}{(2\pi)^4} \, 
\frac{1}{p_E^2} \prod\limits_{j=1}^{N-1} \big( 1 + a_j^2 \, p_E^2 \big)^{-1}
= -\frac{i\lambda}{32 \pi^2} \int_0^\infty \dd u \prod\limits_{j=1}^{N-1} \big( 1 + a_j^2 \, u \big)^{-1} \, ,
\label{eq:goeuclid}
\end{equation}
where we have defined $u \equiv p_E^2$.  The expression on the right of Eq.~(\ref{eq:goeuclid}) can be written in terms of partial
fractions. We assume $N \ge 3$ and find
\begin{equation}
\prod\limits_{j=1}^{N-1} \big( 1 + a_j^2 \, u \big)^{-1} = \sum_{j=1}^{N-1} \frac{d_j}{1+a_j^2 \, u} 
\,\,\,\,\, \mbox{ with } \,\,\,\,\, d_k = \prod\limits_{\substack{j=1\\j\not=k}}^{N-1}  \frac{a_k^2}{a_k^2 - a_j^2}  \,\,\, ,
\end{equation}
which allows easy integration of each term in the sum; this leaves us with
\begin{equation}
-i M^2(k^2) = \frac{i\lambda}{32 \pi^2} \sum_{k=1}^{N-1}  \,d_k \,m_k^2 \ln m_k^2 \, ,\,\,\, \mbox{ where } \,\,\,\,\,  d_k = \prod\limits_{\substack{j=1\\j\not=k}}^{N-1}  \frac{m_j^2}{m_j^2 - m_k^2} \,\, .
\label{eq:finiteNres}
\end{equation}
Here we have expressed the result in terms of the resonance masses, and we have used at an intermediate step that
\begin{equation}
\label{eq:partial-fraction-cancellation}
\sum_{k=1}^{N-1} d_k \,m_k^2 = 0 \,\,\, ,
\end{equation}
which holds for $N \ge 3$. For $N=2$, no such cancellation can occur and one is instead left with a result
that diverges logarithmically with the cutoff.  In the case where $N=3$ 
one finds
\begin{equation}
M^2(k^2) = \frac{\lambda}{32 \pi^2}\, \frac{m_1^2 \, m_2^2}{m_2^2-m_1^2} \, \ln \left(\frac{m_2^2}{m_1^2} \right)\,\,\, .
\label{eq:specialcase}
\end{equation}
Notice that if $m_2 \gg m_1$, the radiative correction to the bare mass squared is positive and with magnitude set by the lightest 
Lee--Wick partner mass $m_1^2$.  The same is true for larger $N$, provided we are far from the asymptotic case considered below; 
when the additional states beyond the first three are made heavy they decouple, leaving the same $N=3$ effective theory, so that the scale of 
the result is again set by $m_1^2$ when $m_2^2$ is also made heavy.

Last, let us note that the general expression in Eq.~(\ref{eq:finiteNres}) is valid for an arbitrary choice of nondegenerate $m_j$.  This will be of use in confirming the results of the next subsection, where we consider the limiting form for the self-energy as $N \rightarrow \infty$ that follows from a specific parametrization of the spectrum.

\subsection{Exactly Solvable Model at Infinite $N$} \label{subsec:b}
\begin{figure}[!htb]
\centering
\includegraphics[width=0.75\textwidth]{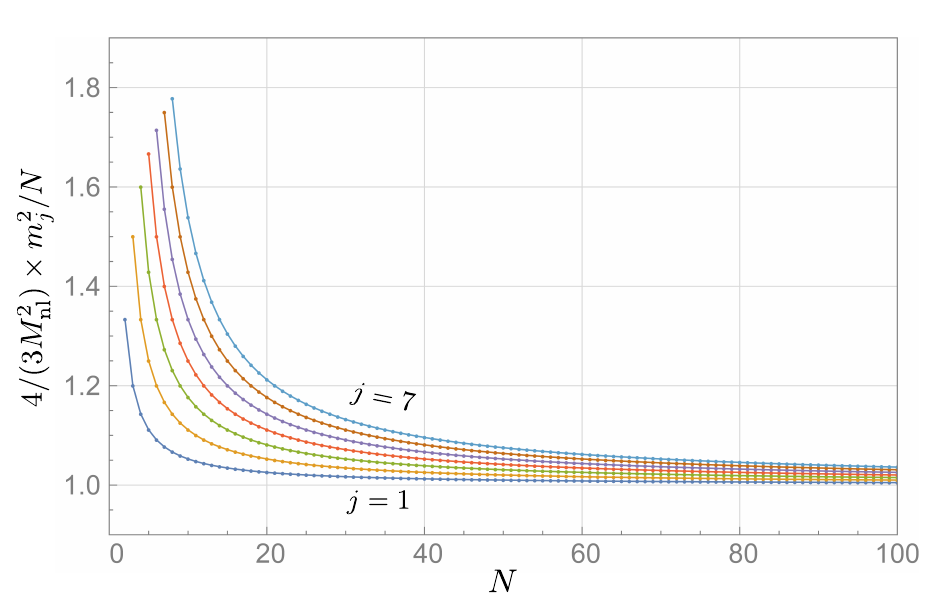}
\caption{We plot the first seven masses for the parametrization \eqref{eq:solvable-model-mass-parametrization}, rescaled by a factor of $N$,
for $j=1$ to $7$ (bottom to top). In the large-$N$ limit, all masses tend to a common large Lee--Wick value (here normalized to unity).}
\label{fig:masses}
\end{figure}

We expect asymptotic nonlocality to be obtained in the theory defined by Eq.~\eqref{eq:hd-lagrangian} if the Lee--Wick masses 
approach a common value as they are simultaneously decoupled and their number is taken to infinity.   Here we implement a concrete ansatz that allows us to achieve this result and to arrive at a closed-form expression for the scalar self-energy.  Consider the parametrization
\begin{align}
a_j^2 = \left( 2 - \frac{j}{N} \right) \frac{2\ell^2}{3N} \, ,
\label{eq:goldenparam}
\end{align}
where $\ell$ denotes the scale of nonlocality. For algebraic convenience we use factors of $N$ rather than $N-1$ in this parametrization of
$a_j$, which reproduces identical large-$N$ asymptotics. For the masses, this parametrization amounts to
\begin{align}
\label{eq:solvable-model-mass-parametrization}
m_j^2 = \frac32 \frac{1}{2 - \frac{j}{N}} N M_\text{nl}^2 \, ,
\end{align}
consistent with Eq.~\eqref{eq:hierarchical-separation}. This parametrization provides for a nondegenerate mass spectrum, for any finite $N$, a
desired feature we discussed earlier.  In Fig.~\ref{fig:masses} we plot the first few mass values $m_j^2$ for different values of $N$.

Equation~(\ref{eq:goldenparam}) allows us to evaluate the product in Eq.~(\ref{eq:goeuclid}) directly.  One can prove the identity
\begin{align}
\prod\limits_{j=1}^{N-1} \left( 1 + a_j^2 \,p_E^2 \right) = \left(\frac{2\ell^2p_E^2}{3N^2}\right)^{N-1} P\left( 1 + N + \frac{3N^2}{2\ell^2p_E^2}, N -1 \right) \, ,
\end{align}
where $P(a,n)$ is the Pochhammer symbol (sometimes also called ``rising factorial'') given by
\begin{align}
P(a,n) \equiv (a)_n \equiv \frac{\Gamma(a+n)}{\Gamma(a)} \, .
\end{align}
In the limiting case of large $N$ we may employ the Stirling identity,
\begin{align}
\Gamma(N+1) \approx \sqrt{2\pi N} \left(\frac{N}{e}\right)^N \, ,
\end{align}
and after some manipulations arrive at 
\begin{align}
\lim\limits_{N\rightarrow\infty} \prod\limits_{j=1}^{N-1} \left( 1 + a_j^2 p_E^2 \right) = e^{\ell^2 p_E^2} \left[ \, 1 - \left(1+\frac{14}{27} \, p_E^2 \ell^2\right)
\frac{p_E^2 \ell^2}{N} + \mathcal{O}\left(\frac{1}{N^2}\right) \, \right] 
\, .
\label{eq:expf}
\end{align}  
The limit point corresponds to a nonlocal exponential propagator, as one could have expected from previous considerations. However, our calculations 
prove that for nondegenerate masses one recovers the exponential propagator as well, which one may regard as a non-trivial consistency check.  
Perhaps more importantly, this calculation demonstrates the emergence of the nonlocal scale as one appropriately decouples an infinite number
of Lee--Wick resonances, {\em i.e.}, how asymptotic nonlocality is achieved.

As a side note, it is well known that analytic continuation in the presence of nonlocal operators of the form $e^{\ell^2\Box}$ is non-trivial due to the existence of essential singularities in the complex momentum plane 
for such entire functions \cite{Pius:2016jsl}. Our construction sidesteps 
these difficulties. For finite $N$ we simply have a Lee--Wick theory in which suitable CLOP-type prescriptions can presumably be used to guarantee 
the proper behavior of the quantum theory \cite{Cutkosky:1969fq}; see also \cite{Anselmi:2017yux}.

At any rate, the subsequent integration over the Euclidean momentum can be performed analytically and one finds the following simple result for the self-energy:
\begin{align}
\label{eq:selfenergy-infinite-N}
\lim\limits_{N\rightarrow\infty} M^2(k^2) = \frac{\lambda}{16\pi^2} \int_0^\infty  \dd p_E \, p_E \,e^{-\ell^2 p_E^2} = \frac{\lambda}{32\pi^2\ell^2} = \frac{\lambda \, M_\text{nl}^2}{32\pi^2} \, .
\end{align}
Note that this result is finite and its scale is set purely by $\ell$, as 
opposed to the finite-and-not-so-large $N$ case where the scale is set by 
the 
lightest Lee--Wick resonance mass.   In other words, the large-$N$ limit has provided us with a hierarchical separation between the scale of the 
lightest Lee--Wick resonance and the potentially much lighter nonlocal scale.  This suggests an interesting approach to evading experimental 
constraints on new particles in more realistic higher-derivative theories 
that are designed to address the hierarchy problem.

\subsection{Large-$N$ Comparison} 
\begin{figure}[!htb]
\centering
\includegraphics[width=0.75\textwidth]{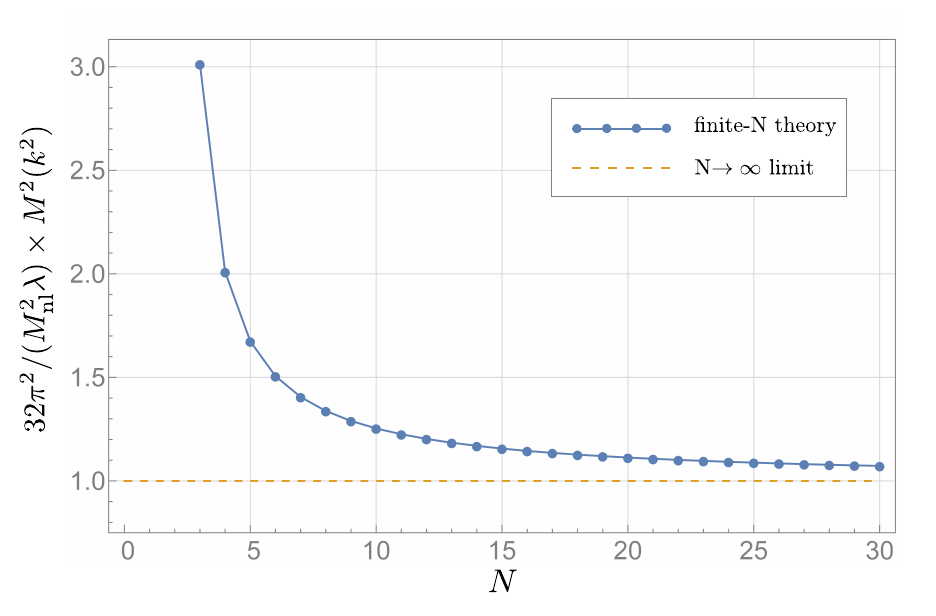}
\caption{For our parametrization \eqref{eq:solvable-model-mass-parametrization} we compare the self-energy at finite $N$ (solid line),
given by Eq.~(\ref{eq:finiteNres}), to the value obtained in the $N\rightarrow\infty$ limit (dashed line), given by Eqs.~(\ref{eq:expf}) and 
(\ref{eq:selfenergy-infinite-N}). It is clear that for large $N$ the results converge.}
\label{fig:self-energy}
\end{figure}

One may wonder whether the different approaches we have presented in Secs.~\ref{subsec:a} and \ref{subsec:b} agree with one another.  In other words, do the $d_k$ prefactors in Eq.~(\ref{eq:finiteNres}), computed in the higher-derivative theory at finite $N$, conspire to give the nonlocal
scale that emerges in Eq.~(\ref{eq:selfenergy-infinite-N}) in the large-$N$ limit?  The analytic calculations would imply that this the case, and upon substituting our mass parametrization \eqref{eq:solvable-model-mass-parametrization} into the exact result \eqref{eq:finiteNres} we can confirm this correspondence numerically as well. This is shown in Fig.~\ref{fig:self-energy}.

Having the general result \eqref{eq:finiteNres} at our disposal, we may also probe different mass parametrizations. For example, the parametrization
\begin{align}
a_j^2 = \left( 1 + \frac{(N-j)^b}{N^b} \right) \frac{b+1}{b+2} \frac{\ell^2}{N} \, , \quad b \in \mathbb{N} \, ,
\end{align}
also reproduces the limit \eqref{eq:selfenergy-infinite-N}.  This supports the notion that (i) our results are independent of how exactly one
chooses to lift the degeneracy in the Lee--Wick spectrum, and (ii) that the relevant scale is indeed the asymptotically nonlocal scale. The only two requirements on the mass functions $m_j$ seem to be nondegeneracy as well as the scaling behavior that $m^2_j\sim N$ as $N\rightarrow\infty$.   
It is interesting to ponder the ultraviolet completions that would lead to higher-derivative theories with similar 
spectra; however, this would likely require an understanding of Planck-scale physics that goes beyond a quantum field theory description.

The numerical results for the one-loop amplitude that we have computed in our local, higher-derivative theory demonstrate that the regulator of loop diagrams in the limit of large but finite $N$ is well approximated by the nonlocal scale that appears in the limiting form of the theory.  We expect that the scale that 
regulates divergences in higher-loop diagrams is no different, based on a dimensional argument:  in the large-$N$ limit, the Lagrangian approaches a 
form where the nonlocal scale is the {\em only} one available to regulate divergences, and no small dimensionless factors appear in the theory.  This implies that our
results should persist in higher-order perturbative calculations, though  we do not consider those in the present work.

\section{Conclusions}\label{sec:conc}
\begin{figure}[!htb]
\centering
\includegraphics[width=\textwidth]{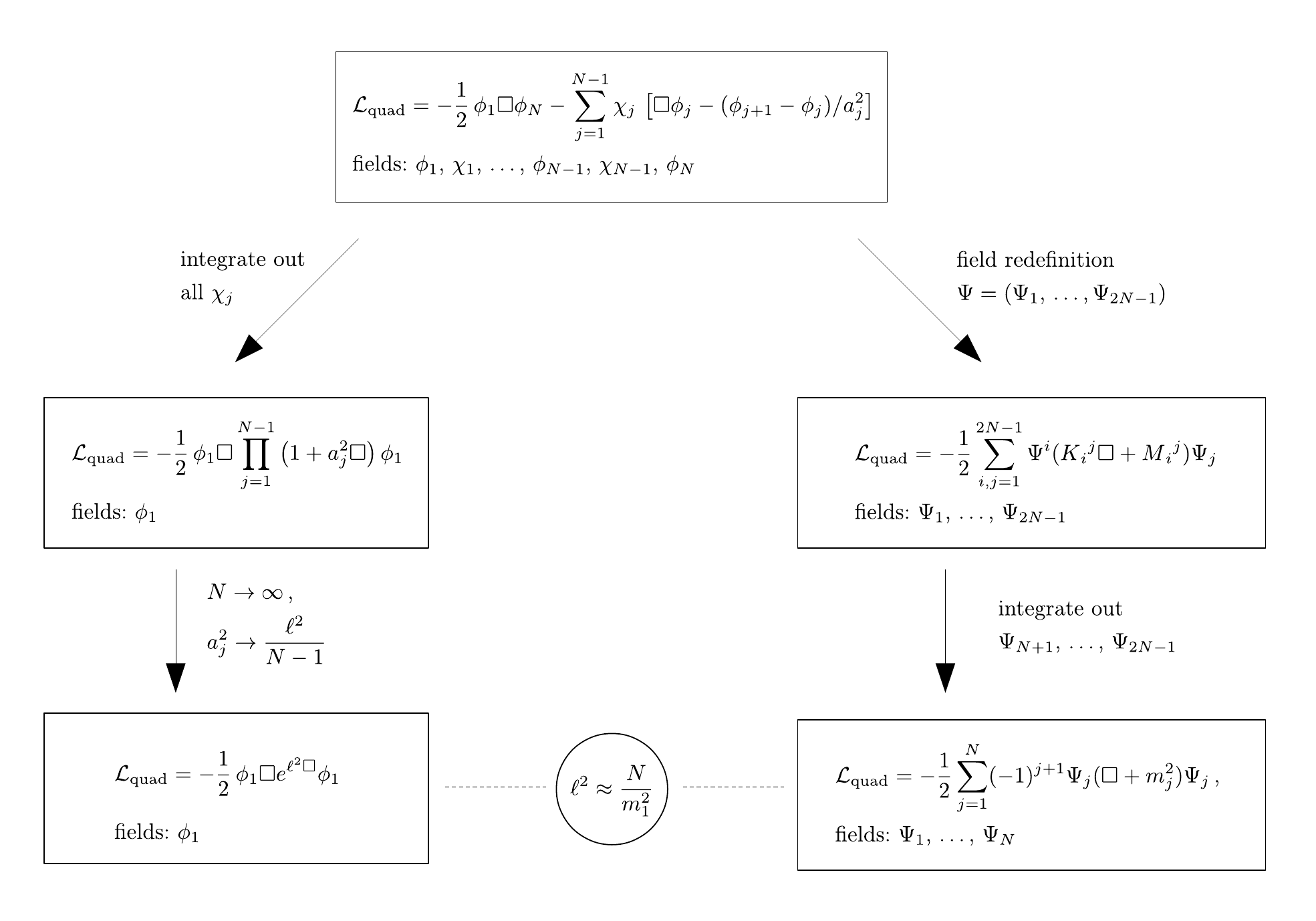}
\caption{A schematic overview of the quadratic terms of the different theories developed in this paper, and how they are interrelated. In the large-$N$ limit, the Lee--Wick resonances tend to infinity, whereas the scale 
of nonlocality $\ell$ remains finite.}
\label{fig:overview}
\end{figure}
Higher-derivative quadratic terms have received substantial attention as a means of obtaining improved 
ultraviolet behavior of loop amplitudes. When these terms are of finite order in the number of derivatives, as in the Lee--Wick standard 
model~\cite{Grinstein:2007mp}, additional poles in the two-point function 
appear that correspond to new particles; these might be 
expected at the TeV scale if they participate in a solution to the hierarchy problem. So far, no such states have been 
observed experimentally. On the other hand, in theories with higher-derivative quadratic terms of infinite order, as in models 
where the $\Box$ operator appears as the argument of an entire function, there are no new particles, but there are other 
complications: the simplest formulation of such nonlocal theories in Minkowski spacetime violate unitarity \cite{Tomboulis:2015esa,Carone:2016eyp,Briscese:2018oyx,Briscese:2021mob,Koshelev:2021orf}, a problem related to 
the fact that there are directions in the complex $p^0$ plane where loop integrands diverge. Another technical complication lies in the fact that
entire functions typically possess essential singularities in the complex 
plane, making a Wick rotation considerably more involved.  
While several ways have been proposed to avoid this problem in nonlocal theories \cite{Pius:2016jsl}, we present a different approach
to the problem in the present work: Formulating a higher-derivative theory with $N$ propagator poles that can be made unitary following 
the same prescriptions as in conventional Lee--Wick theories, it approaches the desired nonlocal form (the exponential of a differential operator)
as $N \rightarrow \infty$. We call this ``asymptotic nonlocality.'' At the same time, the mass spectrum of Lee--Wick partners is decoupled in a
specific way. Even away from the $N \rightarrow \infty$ limit, the effective regulator in loop amplitudes is set by the would-be nonlocal scale,
which can be hierarchically separated from the scale of the lightest Lee--Wick partner. In the present work, we demonstrate this via a simple scalar
model and develop a theory of auxiliary fields that relates the higher-derivative and Lee--Wick forms of the theory; see Fig.~\ref{fig:overview}
for a schematic overview. It is known that the Lee--Wick approach may be generalized to gauge theories, as has been  shown in the Lee--Wick standard
model~\cite{Grinstein:2007mp}, as well as in a higher-derivative extension thereof involving more than one Lee--Wick partner particle~\cite{Carone:2008iw}.
We will study the generalization of the present work to spontaneously broken gauge theories and fields of other spin in a separate publication.

There are other issues that are also worthy of further investigation. The 
form of our theory is nongeneric, as would be expected since the form of
the (well-studied) nonlocal theory that it is designed to match when $N\rightarrow \infty$ is also nongeneric. It would be worthwhile to study the
renormalization of the theory, how it corresponds to the renormalization of the nonlocal theory, and whether this leads to any qualitative changes
in the properties of amplitudes. It would also be interesting to construct the argument for power-counting renormalizability for gauge theories in 
our auxiliary-field construction for arbitrary $N$, similar to what has been achieved for conventional Lee--Wick theories. A generalization to asymptotically
nonlocal gravity would also be of interest.  

The asymptotically nonlocal model presented here provides a convenient interpolation between conventional Lee--Wick theories  and nonlocal quantum 
field
theories, while keeping the best features of both. We look forward to investigating theories of this type in a variety of contexts in future work.

\begin{acknowledgments}  
We thank the NSF for support under Grant PHY-1819575.  
\end{acknowledgments}

\end{document}